\documentclass[twocolumn,showpacs,amsmath, amssymb, prd]{revtex4-1}
\usepackage[pdftex]{graphicx}% Include figure files
\usepackage{dcolumn}% Align table columns on decimal point
\usepackage{bm}% bold math
\usepackage{color}% color
\usepackage{mathrsfs}

\newcommand{\msun}{{\rm M}_\odot}
\newcommand{\zsun}{{\rm Z}_\odot}
\newcommand{\pc}{{\rm pc}}
\newcommand{\kpc}{{\rm kpc}}
\newcommand{\mpc}{{\rm Mpc}}

\newcommand{\kms}{{\rm km~s}^{-1}}
\newcommand{\gpcyr}{{\rm Gpc}^{-3}~{\rm yr}^{-1}}

\newcommand{\apjl}{Astrophys. J.}
\newcommand{\apjs}{ApJS}
\newcommand{\aap}{A$\&$A}

\newcommand{\araa}{ARA$\&$A}

\newcommand{\mnras}{Mon. Not. R. Astron. Soc.}

\newcommand{\aj}{AJ}

\graphicspath{{figures/}}

\begin{document}

\title{Probing stellar binary black hole formation in galactic nuclei via the imprint\\
of their center of mass acceleration on their gravitational wave signal}
\author{Kohei Inayoshi$^{1}$}
\author{Nicola Tamanini$^{2}$}
\author{Chiara Caprini$^{3}$}
\author{Zolt\'an Haiman$^{1,4}$}
\affiliation{
${}^{1}$Department of Astronomy, Columbia University, 550 W. 120th St., New York, NY, 10027, USA\\
${}^{2}$ Institut de Physique Th\'eorique, CEA-Saclay, CNRS UMR 3681, Universit\'e Paris-Saclay, F-91191 Gif-sur-Yvette, France\\
  ${}^{3}$ Laboratoire Astroparticule et Cosmologie, CNRS UMR 7164, Universit\'e Paris-Diderot, 10 rue Alice Domon et L\'eonie Duquet, 75013 Paris, France\\
  $^{4}$Department of Physics, New York University, New York, NY 10003, USA
}

\begin{abstract}
Multi-frequency gravitational wave (GW) observations are useful probes
of the formation processes of coalescing stellar-mass binary black
holes (BBHs).  We discuss the phase drift in the GW inspiral waveform
of the merging BBH caused by its center-of-mass acceleration.  The
acceleration strongly depends on the location where a BBH forms within
a galaxy, allowing observations of the early inspiral phase of
LIGO-like BBH mergers by the Laser Interferometer Space Antenna (LISA)
to test the formation mechanism.  In particular, BBHs formed in dense
nuclear star clusters or via compact accretion disks around a nuclear
supermassive black hole in active galactic nuclei would suffer
strong acceleration, and produce large phase drifts measurable by
LISA.  The host galaxies of the coalescing BBHs in these scenarios can
also be uniquely identified in the LISA error volume, without
electromagnetic counterparts.  A non-detection of phase drifts would
rule out or constrain the contribution of the nuclear formation
channels to the stellar-mass BBH population.
\end{abstract}

\date{\today \hspace{0.2truecm}}

\maketitle

\section{Introduction}

Advanced LIGO has so far detected gravitational waves from three stellar
binary black hole (BBH) mergers 
\cite{Abbott_PRL_2016,Abbott_PRL_2_2016,LIGO_BBH,LIGO_O2_2017}.  Several scenarios
for the origin of such massive compact BBHs have been proposed
\cite{Abbott_2016_Astro}, through the evolution of isolated massive
stellar binaries
\citep{Belczynski_2004,Dominik_2012,K14,Belczynski_2016_nature,Inayoshi_2016},
dynamical formation in dense stellar clusters
\cite{PortegiesZwart_2000,O'Leary_2009, Rodriguez_2016,Antonini_2016} or 
in active galactic nuclei (AGN)
\cite{Bellovary+2016,Bartos_2016,Stone_2017}.

Recently, Ref.~\cite{Sesana_2016} pointed out that the early
inspiral of GW150914-like BBHs can be measured by the
space-based detector LISA \cite{Audley:2017drz}. The BBH coalescence
rate inferred from the LIGO detections implies that $\sim 10-100$ of
such BBHs will be individually resolved by LISA and then merge in the
LIGO band in $\leq 10$ yr.  
These LISA observations alone can determine the coalescence time 
with an accuracy of $\sim 10$ s and the sky position to within $< 1$ deg$^2$, 
allowing advance planning of electromagnetic (EM) observations of the merger.

Multi-frequency GW observations by LISA and LIGO are also useful to
distinguish formation scenarios of stellar-mass BBHs
(e.g., measurements of spin-orbit misalignments by LIGO 
\cite{Rodriguez_2016b,Vitale_2017,Stevenson_2017}).  
LISA could detect non-zero eccentricities of the merging BBHs \cite{Seto_2016, Nishizawa_2016,Breivik_2016}.  
Measurable eccentricities are expected in formation channels involving dynamical
interactions in dense stellar clusters \cite{O'Leary_2009}, or as a
result of the interaction with AGN accretion disks
\cite{Bellovary+2016,Bartos_2016,Stone_2017}, but not if typical BBHs
form via isolated binaries \cite{Belczynski_2016_nature}.
The predicted eccentricities are uncertain and alternative
ways to distinguish formation channels remain useful.

Another important advantage of low-frequency GW observations by LISA
(and/or by DECIGO \cite{DECIGO_2006}) is that the acceleration of the
merging BBHs can produce a measurable phase drift in their GW inspiral waveform.  
In the cosmological context, the apparent acceleration 
caused by the time evolution of the Hubble expansion is weak, 
but if detected, it would allow us to measure the accelerated expansion directly 
\cite{Seto_2001,Nishizawa_2012}.
However, the peculiar acceleration of the coalescing BBHs due to astrophysical 
processes could be much larger than that produced by the cosmic expansion
\cite{Yunes_2011b,Bonvin_2016}.

In this paper, we discuss the possibility to distinguish the
formation channels of merging BBHs, focusing on the binary motion
inside the host galaxy.  We show that BBHs located in dense nuclear
star clusters or in compact accretion disks around a nuclear supermassive BH (SMBH)
suffer strong acceleration, and produce a large phase drift measurable by LISA 
\footnote{For a stellar-mass BBH located exceptionally close
($\lesssim 10^{11}$cm) to an SMBH, the acceleration,
together with other relativistic corrections to the GW waveform,
could be measured by LIGO alone~\cite{Meiron_2017}; see
\S~IV.}.
Since the acceleration effect strongly depends on the location where
BBHs form within a galaxy, observations by LISA of stellar-mass BBH
mergers offer a test of proposed formation scenarios.

\vspace{-0.5\baselineskip}
\section{Phase drift in gravitational waves}
\vspace{-0.5\baselineskip}

We consider a coalescing BBH with redshifted chirp mass 
$M_{\rm cz}$ ($M_{\rm cz,40}\equiv M_{\rm cz}/40 M_\odot$) 
in the LISA band with a signal-to-noise
ratio (SNR) of $\rho>8$ which will merge in the advanced LIGO/Virgo band 
in $\tau_{c}\lesssim10$ yr~\cite{Sesana_2016}. The rest-frame GW
frequency of the coalescing BBH is given by (see e.g.~\cite{Peters_1964,michele})
\begin{equation}
f\simeq 13.8~M_{\rm cz,40}^{-5/8}\left(\frac{\tau_{c}}{4~{\rm yr}}\right)^{-3/8}~{\rm mHz}\,.
\end{equation}
The SNR accumulated during the LISA observation time $\delta t$ 
can be approximated as \cite{Seto_2016}
\begin{align}
\rho
\simeq 51~d_{L,100}^{-1}~
M_{\rm cz,40}^{5/3}~
f_{14}^{2/3}~
\delta t_{5}^{1/2},
\label{eq:snr}
\end{align}
where $d_{L,100}\equiv d_{L}/(100~\mpc)$ is the luminosity distance to the GW source and
$f_{14}\equiv f/(14~{\rm mHz})$ is a representative frequency.
Throughout the paper we assume a LISA configuration with six links, 2 million km arm length and 
mission duration of $\delta t=5$ yr $(\delta t_5 \equiv \delta t/5~{\rm yr})$. 
LISA's noise curve $S_{\rm n}(f)$ has been taken from \cite{Klein_2016}. 
In Eq.~\eqref{eq:snr}, we have set  LISA's sensitivity to
$\sqrt{S_{\rm n}(f)}=7\times 10^{-21}/\sqrt{\rm Hz}$
at $f\simeq 10$ mHz.  

From the BBH merger rate inferred from the existing LIGO events,
$R\simeq 9-240~\gpcyr$ \cite{LIGO_BBH}, the space density of BBHs
inspiralling near $f\simeq 10^{-2}$ Hz at a given time can be
estimated as
\begin{align}
n_{\rm m}\equiv R\,\frac{f}{\dot{f}}=10^{-6}
~f_{14}^{-8/3}~
M_{\rm cz,40}^{-5/3}~
R_{100}
~\mpc^{-3},
\label{eq:nm}
\end{align}
where $R_{100}=R/(100~\gpcyr)$. 
Thus, below we require $60~\mpc <d_L<640~\mpc$, in order to ensure
that at least one event ($N_{\rm m}\equiv 4\pi n_{\rm m} d_L^3/3>1$,
evaluated with $R_{100}=1$) occurs in the local cosmic volume with
a total SNR $\rho >8$, during a $\delta t=$ 5yr LISA mission lifetime.

Next, we consider the impact of the center-of-mass (CoM) acceleration
of a merging BBH. Over $\delta t$, the source will appear to change
its redshift by an amount
\begin{equation}
\delta z_{\rm acc}\simeq \frac{a_{\rm CoM}\delta t}{c} 
\equiv 1.7\times 10^{-7}
~\delta t_5~\left(\frac{\epsilon}{10^4}\right),
\end{equation}
where we have expressed the acceleration $a_{\rm CoM}=v^2_{\rm acc}/r$
along the line of sight
in terms of a characteristic velocity $v_{\rm acc}$ and distance $r$
(interpreted below as the orbital velocity and distance from the
barycenter of the host galaxy), and defined the dimensionless
acceleration parameter $\epsilon$
\begin{align}
\epsilon \equiv 10^4 \left(\frac{v_{\rm acc}}{100~\kms}\right)^2
\left(\frac{r}{1~\pc}\right)^{-1} . 
\end{align}
We define a variable $Y$ which accounts for the CoM acceleration of a
merging BBH by
\begin{align}
Y \equiv \frac{1}{2(1+z)}\cdot \frac{\delta z_{\rm acc}}{\delta t}
\simeq 1.5\times 10^{-8}~\epsilon_{\rm z,4}~{\rm yr}^{-1},
\label{eq:YH}
\end{align}
where $\epsilon/(1+z)\equiv 10^4\epsilon_{\rm z,4}$~\footnote{
Note that the redshift drift due to the cosmic expansion is negligible
for the range of $\epsilon$ we consider here.}.  
The CoM acceleration causes a linear frequency drift $\delta f \propto Y\delta t$ 
and the corresponding phase drift in the GW inspiral waveform is expressed
as \cite{Bonvin_2016}
\begin{align}
\delta \Phi _{\rm acc}&
\simeq 1.0~\epsilon_{\rm z,4}~
\delta t_5~
M_{\rm cz,40}^{-5/3}~
f_{14}^{-5/3}.
\label{Phiacc}
\end{align}
The total number of GW cycles without the acceleration is $\mathcal{O}(10^6)$
for stellar-mass binaries, and Eq.~\eqref{Phiacc} gives about one extra cycle for the reference values of the parameters.

%%%%%%
%	Fig.1  %
%%%%%%
\begin{figure}
\includegraphics[width=90mm]{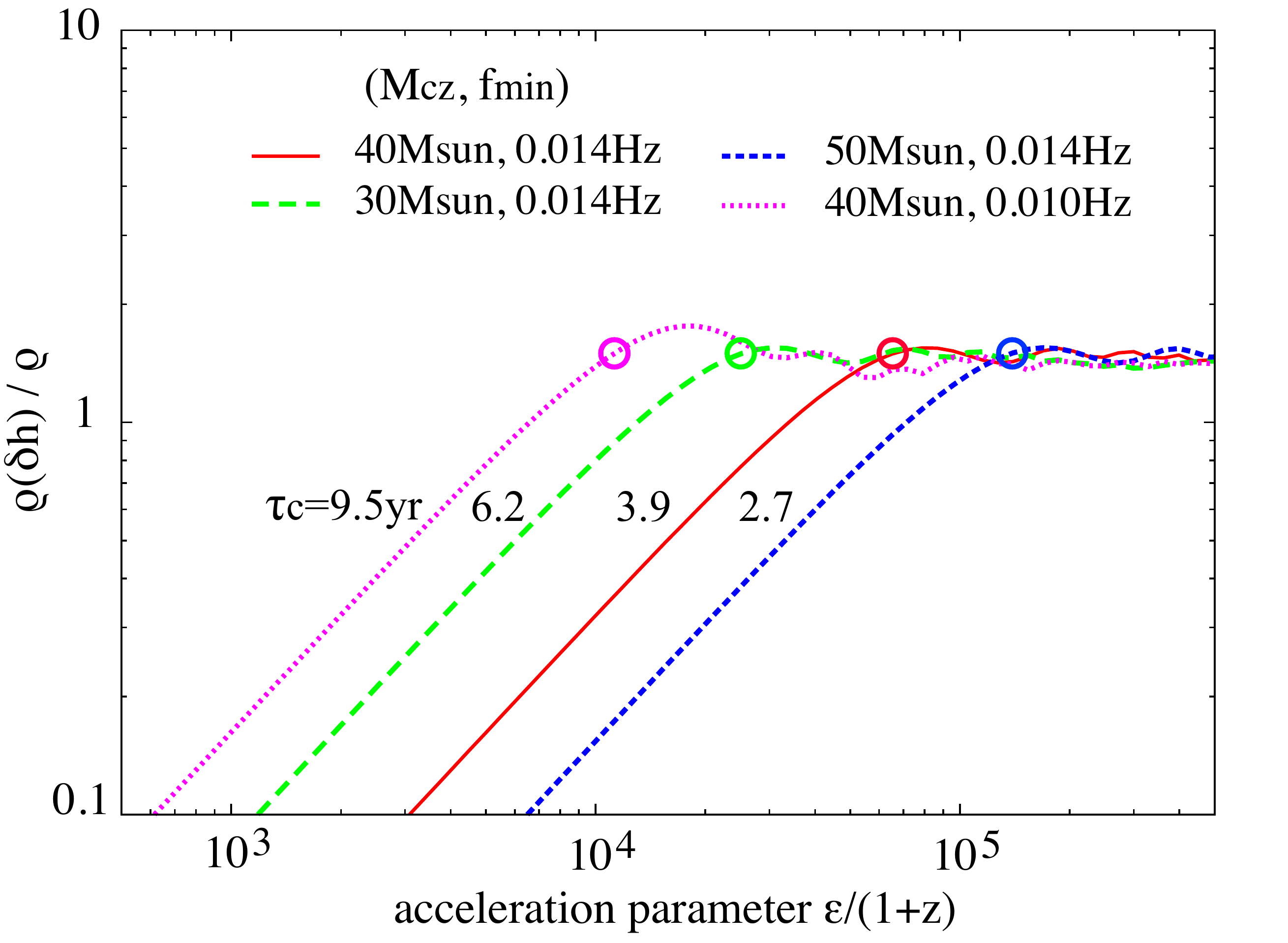}
\caption{The relative SNR of the deviation in the GW inspiral waveform
  caused by the CoM acceleration $\epsilon$, for different
  combinations of the redshifted chirp mass $M_{\rm cz}$ and the
  frequency $f_{\rm min}$ when the LISA observation begins.  The
  corresponding time to coalescence $\tau_{\rm c}$ is indicated in the
  figure. The LISA observation is limited to the duration $\delta t=5$
  yr.  The relative SNRs saturate at $\epsilon>\epsilon_{\rm crit}$ ($\delta
  \Psi_{\rm acc}\gtrsim 1$) shown by open circles.}
\label{fig:SNR}
\end{figure}

%%%%%%
%	Fig.2  %
%%%%%%
\begin{figure}
\centering\includegraphics[width=87mm]
{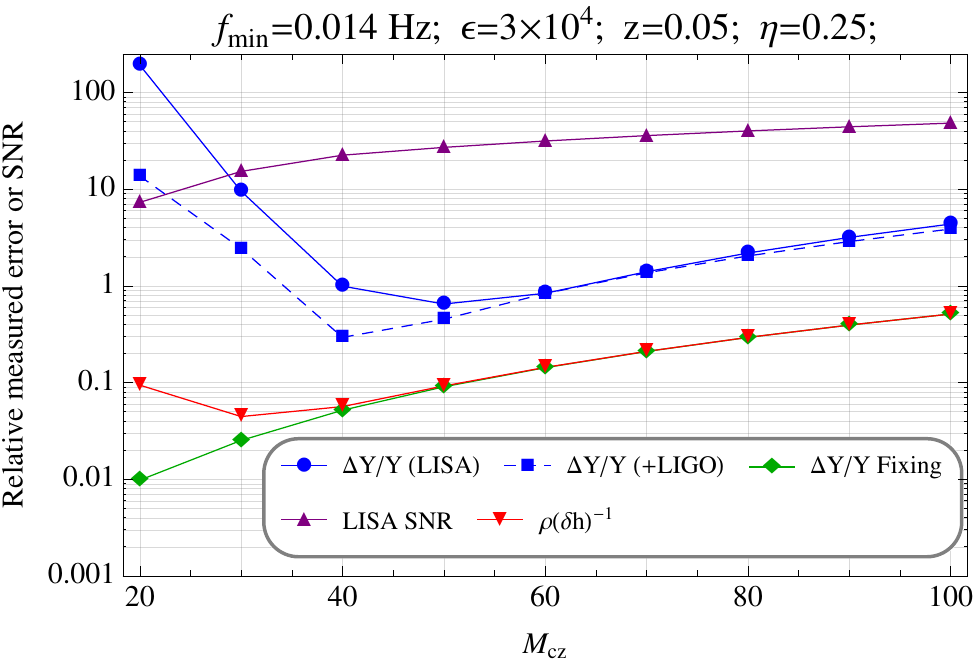}
\caption{The 1$\sigma$ errors $\Delta Y/Y$ from LISA alone (blue solid) and
  LISA + LIGO (i.e. assuming that the coalescence time $t_c$ has
  been fixed by LIGO; blue dashed).  For our fiducial case ($f_{\rm
    min}=0.014$ Hz, $\epsilon =3\times 10^4$, $z=0.05$ and
  $\eta=0.25$), non-zero acceleration can be detected, i.e., $\Delta
  Y/Y<1$, for merging BBHs with $35~\msun\lesssim M_{\rm cz}\lesssim
  63~\msun$.  The LISA SNR (purple) and the inverse of the numerically
  computed $\rho(\delta h)$ due to the CoM acceleration (red) are
  shown.
  The green curve shows the Fisher error assuming all parameters
  but $Y$ are fixed: 
  for $M_{\rm cz}\gtrsim 35~\msun$ it coincides with the inverse of  
  $\rho(\delta h)$, validating the Fisher analysis; for lower $M_{\rm cz}$ 
  the Fisher analysis does not capture the saturation effect discussed 
  below Eq.~\eqref{rhodeh} and shown in Fig.~\ref{fig:SNR}.
    }
\label{fig:comparison}
\end{figure}

To detect the phase drift in the GW inspiral waveform, the strain
perturbation $\delta h(f)=h(f)[1-e^{i \delta \Psi_{\rm acc}(f)}]$ 
must have a significant SNR \cite{Yunes_2011,Kocsis_2011}, where 
$\delta \Psi_{\rm acc}(f)$ is the phase drift in frequency space
\cite{Cutler_Flanagan_1994}
\begin{align}
\delta \Psi_{\rm acc}(f)\simeq -0.59~\epsilon_{\rm z,4}~
M_{\rm cz,40}^{-10/3}~
f_{14}^{-13/3}.
\end{align}
Fig.~\ref{fig:SNR} shows the relative SNR of the perturbation
$\rho(\delta h)/\rho =\left[\int^{f_{\rm max}}_{f_{\rm min}}df\,\frac{|\delta h(f)|^2}{S_n(f)} 
/ \int^{f_{\rm max}}_{f_{\rm min}}df\,\frac{|h(f)|^2}{S_n(f)}\right]^{1/2}$ for
different combinations of the redshifted chirp mass $M_{\rm cz}$ and
the frequency $f_{\rm min}$ when the observation begins ($f_{\rm max}$
is the smaller between twice the inner-most-stable-circular-orbit
frequency or the frequency reached in $\delta t=5$yr).

For small $\epsilon$, the SNR of the perturbation
is proportional to $|\delta \Psi_{\rm acc}|$ \cite{Kocsis_2011, Hayasaki_2013} 
and is given by
\begin{align}
\rho(\delta h)
\simeq 16~\epsilon_{\rm z,4}~d_{L,100}^{-1}~
\delta t_5^{1/2}~
M_{\rm cz,40}^{-5/3}~
f_{14}^{-11/3}.
\label{rhodeh}
\end{align}
For larger $\epsilon(>\epsilon_{\rm crit})$, when the phase drift approaches a
full cycle, the relative SNR of $\rho (\delta h)/\rho$ saturates at a roughly constant value ($\simeq 1.5$) because of de-phasing.
Computing the relative SNR numerically, we found the critical
accelerations to be $\epsilon_{\rm crit}/(1+z)\simeq 6.6\times 10^4~M_{\rm cz,40}^{10/3}~f_{14}^{13/3}$.

%%%%%%
%	Fig.3  %
%%%%%%
\begin{figure*}
\centering\includegraphics[width=160mm]
{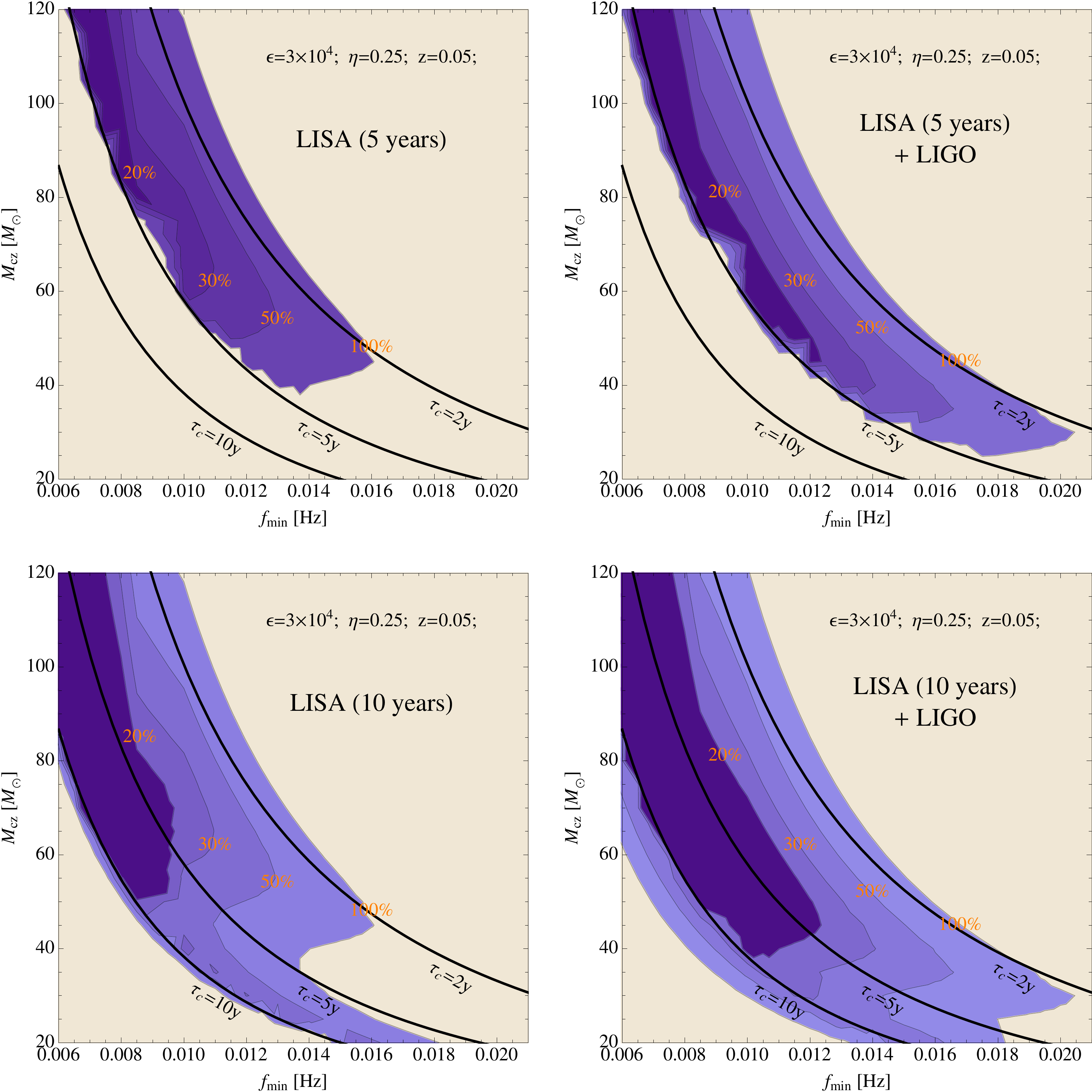}
\caption{Contours of the marginalized $1 \sigma$ error $\Delta Y / Y$
  in the $f_{\rm min}$--$M_{\rm cz}$ parameter space, provided by
  LISA alone (top left panel) and LISA + LIGO (top right panel) assuming a $5$ yr mission, 
  and LISA alone (bottom left panel) and LISA + LIGO (bottom right panel) assuming a $10$ yr mission.  
  The solid curves indicate constant times to coalescence.  Merging BBHs with 
  2 yr $\lesssim \tau_c \lesssim$ 5 (or 10) yr provide the best combinations
  of $f_{\rm min}$ and $M_{\rm cz}$ to probe the CoM acceleration.}
\label{fig:Mc_fmin_contourplots}
\end{figure*}

In order to estimate the LISA error on the acceleration parameter $Y$,
including possible degeneracies with other system parameters, we
perform a Fisher matrix analysis.  We adopt the six parameters
$M_{\rm cz}$, $\Phi_c$, $t_c$, $\eta$, $d_L$ and $Y$, where $\Phi_c$
is the phase at the coalescence time $t_c$, and $\eta$ is
the symmetric mass ratio.  We further follow Ref.~\cite{Bonvin_2016}
and adopt the sky-averaged GW waveform of $h(f) =A(f) \,
\exp[i\Psi(f)]$ with the amplitude $A(f)$ at Newtonian order, and the
phase $\Psi(f)$ at 3.5PN order, plus the contribution of the CoM
acceleration effect:
\begin{align}
\label{Psi3.5PN}
\Psi(f) = 2\pi f t_c -\frac{\pi}{4} -\Phi_c + \Psi_{\rm PN}(f,\eta) + \delta \Psi_{\rm acc}(f).
\end{align}
The explicit form of $\Psi_{\rm PN}(f)$ is given
in~\cite{Mishra:2010tp}.

Fig.~\ref{fig:comparison} shows the marginalized $1 \sigma$ error
$\Delta Y/Y$ provided by LISA alone (solid blue) for our fiducial
values of $\epsilon=3\times 10^4$, $f_{\rm min}=0.014\, {\rm Hz}$,
$z=0.05$ ($d_L\simeq 200~\mpc$; well inside the horizon of both LIGO
and LISA)
and $\eta=0.25$.  The error is small enough to detect non-zero $Y$
(i.e., $\Delta Y/Y<1$) at $35\lesssim M_{\rm cz}/\msun \lesssim 63$
because the GW chirping helps break the degeneracies among the
waveform parameters.  
When the binary frequency hardly evolves during
the LISA observation, i.e., for lower $M_{\rm cz}$ (and/or $f_{\rm min}$), 
strong degeneracies remain and render the acceleration undetectable.
For higher masses (and/or $f_{\rm min}$), the binary
exits the LISA band more rapidly, diminishing the SNR.

LIGO observations during/after the LISA operation time can reduce
parameter degeneracies, by detecting the merger event and fixing the
coalescence time $t_c$.  As shown by the dashed blue curve in
Fig.~\ref{fig:comparison}, for low masses the error $\Delta Y/Y$ provided by LISA +
LIGO (i.e. $t_c$ fixed) is reduced by a factor of $3-10$ from that by LISA alone.  
As a result, the best error estimate in this case is given by 
$\Delta Y/Y\simeq 0.3$ at $M_{\rm cz}=40~\msun$.

In Fig.~\ref{fig:Mc_fmin_contourplots}, we show the measurement error
$\Delta Y/Y$ as a function of $M_{\rm cz}$ and $f_{\rm min}$. 
In this case we also present the results for a LISA mission lasting 10 yr to show 
how they could improve: the nominal mission duration is 4 yr but a duration in flight 
up to 10 yr is conceivable \cite{Audley:2017drz}. 
In a suitable range of values for $f_{\rm min}$, for a 5 yr LISA mission without any input from LIGO (top left), the acceleration effect can be detected with an
error of $\Delta Y/Y<0.5(1)$ only for relatively massive binaries with
$M_{\rm cz}\gtrsim50(40)~\msun$.  Fixing $t_c$ with LIGO (top right) extends
the same limits down to $M_{\rm cz}\gtrsim 35(25)~\msun$.
A 10 yr LISA mission instead can provide errors of $\Delta Y/Y<0.5$ for masses $M_{\rm cz}\gtrsim 25~\msun$, even without detecting the merger with LIGO (bottom left), while for BBHs with higher masses the acceleration effect can be measured with more accuracy: $\Delta Y/Y<0.2$ for $M_{\rm cz}\gtrsim 50~\msun$.
Fixing the merger time with LIGO (bottom right) improves these results yielding the possibility of detecting the acceleration effect with $\Delta Y/Y<0.5$ for $M_{\rm cz}$ below $20~\msun$, while $\Delta Y/Y<0.2$ can be reached for $M_{\rm cz}\gtrsim 40~\msun$.
Long-lived binaries with $\tau_c>$ 5 (10) yr do not chirp rapidly enough to break
parameter degeneracies, whereas short-lived binaries with $\tau_c<2$
yr do not spend sufficient time in the LISA band to accumulate SNR.
We conclude that merging BBHs with 2 yr $\lesssim \tau_c \lesssim$ 5 (10)
yr provide the best combinations of $f_{\rm min}$ and $M_{\rm cz}$ to
probe CoM acceleration.

\begin{figure}
\centering\includegraphics[width=85mm]
{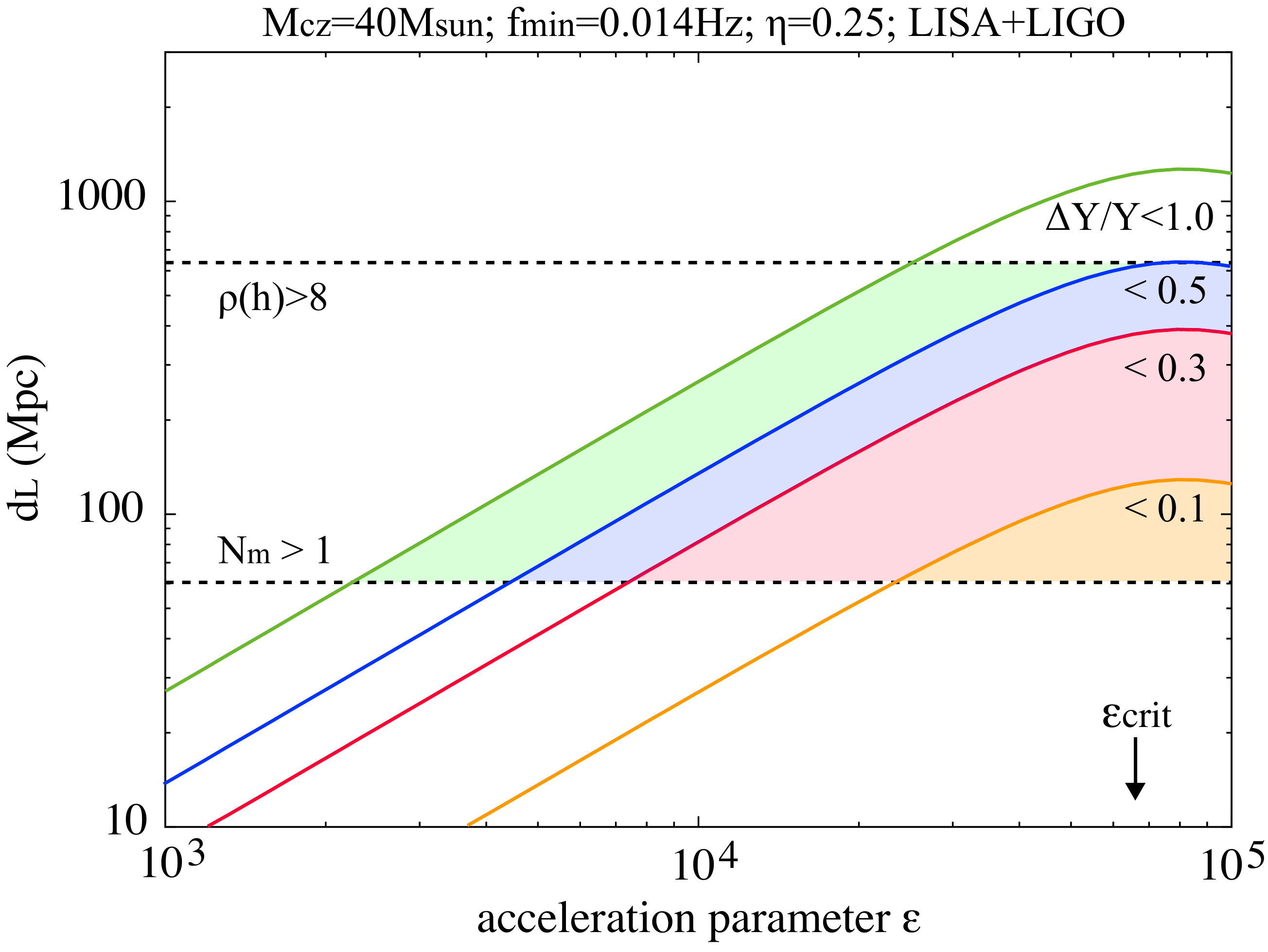}
\caption{GW phase drift detection conditions in the $\epsilon$--$d_L$
parameter space.  The four solid curves mark marginalized 1$\sigma$
errors $\Delta Y/Y<0.1$ (orange), $<0.3$ (red), $<0.5$ (blue) and $<1.0$ (green).
The two horizontal dotted lines show a maximum distance ($D <
640~\mpc$) for a total SNR $\rho(h)\geq 8$ and a minimum distance
($D > 60~\mpc$) to find a BBH at $f\simeq 0.014$ Hz
(i.e., $N_{\rm m}\equiv 4\pi n_{\rm m}d_L^3/3>1$).  
The maximum distance saturates and the Fisher analysis is invalid for large
accelerations $\epsilon>\epsilon_{\rm crit}$, marked by an arrow
(see below Eq. \ref{rhodeh}).  }
\label{dL_vs_eps}
\end{figure}

Fig.~\ref{dL_vs_eps} presents the maximum distances out to which phase
drifts can be measured with errors of $\Delta Y/Y<0.1-1.0$.  
We consider equal-mass BBH mergers with
$M_{cz}=40~\msun$ and $f_{\rm min}=0.014$ Hz.  The horizontal dashed
lines show the conditions $60~\mpc <d_L<640~\mpc$ discussed below
Eq. (\ref{eq:nm}).  
For $\epsilon>\epsilon_{\rm crit}$, 
the maximum distance does not increase linearly with $\epsilon$ 
because of the saturation of the relative SNR discussed below Eq. (\ref{rhodeh}) 
and shown in Fig.~\ref{fig:SNR}.
For merging BBHs located in the shaded region, the phase drifts in the GW inspiral
waveform can be observed.

%%%%%%
%	Fig.4  %
%%%%%%

\begin{table*}[t]
\label{Scenarios}
\centering
\begin{tabular}{ c || c | c | c | c | c | c | c}
\hline
  scenario & ~$v~(\kms)$~ & ~~~$r~(\pc)$~~~ & ~~~~~~$\epsilon$~~~~~~ 
  & ~~$d_{L, {\rm obs}}~(\mpc)$~~ & ~$n_{\rm host}~(\mpc^{-3})$~ & ~~$n_{\rm host}V_{\rm eff}$~~ & ~~$n_{\rm m}V_{\rm eff}$~~\\
  \hline                        
 {\it Field binaries (A)} & & & & & &\\
  formed at $z\simeq 0$ & $\sim 200$ & $>5\times 10^3$ & $<10$ & $\sim 0.2$ & $\sim 2\times 10^{-2}$ \cite{2008ApJ...675.1459K}& $\ll 1$ & $\ll 1$\\
  formed at $z\simeq 3$ & $\sim 300$ & $10^3-10^4$ & $10-100$ & $0.2-2$ & $\sim 5\times 10^{-4}$ \cite{Bell_2003}& $\lesssim 0.02$ & $\ll 1$\\
  \hline
 {\it Dense stellar systems (B)}& & & &&& \\
  globular clusters & $\sim 200$ & $\sim 5\times 10^4$ & $\sim 1$ & $\sim 0.02$ & $\sim 1$ \cite{Harris_2013} & $\ll 1$ & $\ll1$\\
  {\bf nuclear star clusters} & $30-100$ & $\sim 1$ & {\boldmath $10^3-10^4$} & $20-200$ & $\sim 0.01$ \cite{Miller_2009}& $\lesssim 3\times 10^5$ & $\lesssim 30$\\
  \hline
  {\it AGN disks (C)} &&&&&&\\
    {\bf formed in disk} & $\sim 200$ & $\sim 1$ & {\boldmath $\sim 10^4$} & $\sim 200$ & $\sim 10^{-5}$ \cite{Ueda_2014}& $\sim 300$ & $\sim 30$\\
    {\bf captured or migrated in}  & $\sim 600$ & $\sim 0.1$ & {\boldmath$\sim 10^5$} & $\sim 950$ & $\sim 10^{-5}$~~~~~ & $\sim 10^4$ & $\sim 10^3$\\
  \hline
 {\it Very high-redshift (D)} & & & & &&\\
  Population III & $\sim 200$ & $\lesssim 10^3$ & $10-100$ & $0.2-2$ & $\sim 2\times 10^{-2}$ \cite{2008ApJ...675.1459K}& $\lesssim 0.7$ & $\ll 1$\\
  \hline
\end{tabular}
\caption{The first four columns show, in each BBH formation scenario,
  the expected center-of-mass orbital velocity ($v$) and radius ($r$),
  and the acceleration parameter ($\epsilon$).  Col. 5 shows the
  maximum distance ($d_{L, {\rm obs}}$) at which the phase drift can
  be measured by LISA with an SNR of $\rho(\delta h)>8$, corresponding
  to $\Delta Y/Y<0.5$.  In Col. 6, the number densities of the host
  objects are shown.  Cols. 7 and 8 show the number of host objects
  ($n_{\rm host}$) and of GW events ($n_{\rm m}$) in the LISA band,
  within the cosmic volume of $V_{\rm eff}=4\pi d_{\rm eff}^3/3$,
  where $d_{\rm eff}={\rm min}(d_{L, {\rm obs}},640~\mpc)$ and $n_{\rm
    m}=Rf/\dot{f}\simeq 10^{-6}~\mpc^3$.  Here we adopt our fiducial
  case: $M_{\rm cz}=40~\msun$, $f_{\rm min}=0.014$ Hz, and
  $\eta=0.25$.  In the three scenarios indicated by boldface, the acceleration
  is large and measurable.  }
\end{table*}

\vspace{-0.5\baselineskip}
\section{Formation scenarios of LIGO BBHs and corresponding acceleration}
\vspace{-0.5\baselineskip} 

In this section, we review proposed stellar-mass BBH formation
scenarios, from field binaries (A), dynamical formation in dense
stellar systems (B) and in AGN accretion disks (C) and massive
high-redshift binaries (D). We discuss the typical value of the
acceleration parameter $\epsilon$ expected in each case, summarized in
Table~1.

\vspace{1mm} \emph{A. Field binaries} --- A compact ($\lesssim 0.1$
AU) massive stellar binary could form a BH remnant coalescing due to
GW emission in a Hubble time.  Such BBHs are formed in low-metallicity
star forming regions \cite{Abbott_2016_Astro}, possibly over an
extended range of redshifts ($0\lesssim z\lesssim 3$;
e.g. Ref.~\cite{Dominik_2012}).

In the nearby universe, most star-formation occurs in disks of spiral
galaxies, within their half-light radii of $\sim 5$ kpc
\cite{vanderKruitSearle1982}.  Assuming that the stars are orbiting
around the center of the galaxy at the circular velocity $\sim
200~\kms$ of a typical disk galaxy, the acceleration parameter is
$\epsilon \simeq 8$~\footnote{The line-of-sight acceleration will be 
reduced by $\cos \theta $ if the instantaneous acceleration vector is inclined by an
angle $\theta$ to the LOS; see discussion.}.  However, LIGO BBHs
are expected to arise from massive stellar binaries with
$Z<0.1~\zsun$~\cite{Abbott_2016_Astro}.  Since metallicities decrease
farther out in the disk \cite{Pilkington_2012}, BBH formation could
occur preferentially at these larger radii, where the acceleration
parameter is reduced to $\epsilon \sim O(1)$.

A large fraction of low-metallicity massive (binary) stars could form
in high-redshift star-forming galaxies.  Their host galaxies will
undergo several mergers and most of these binaries may end up in the
cores of massive elliptical galaxies.  These old remnant BBHs would be
located in the core with a typical size of a few kpc
\cite{Faber_1997,Yu_2002} and with the circular velocity of $v \sim
300~\kms$ \citep{Kormendy_Ho_2013}, resulting in somewhat larger
accelerations of $\epsilon \simeq 10-100$.

\vspace{1mm}
\emph{B. Dynamical formation in dense stellar systems} ---
Two single BHs can be paired when they interact and form a bound
binary in a dense stellar system, either through a chance close
fly-by, or involving a third object.  These processes likely occur in
globular clusters (GCs), nuclear star clusters (NSCs) and around
SMBHs in galactic nuclei.

Most GCs are in orbit inside dark matter (DM) halos with $M \simeq
10^{12}~\msun$, because galaxies in such a mass range contain most of the 
present-day stellar mass, and the number of GCs scales with their host galaxy mass
\cite{Harris_2013}.  The acceleration parameter is as low as $\epsilon
\lesssim 1$, for the circular velocity of $v \simeq 200~\kms$ at $r
\sim 50~\kpc$ ($\sim$ half of the virial radius).  On the other hand,
the BBHs also orbit inside GCs, where the velocity dispersion is at
most $\sim 10~\kms$ and half-light radii are $\sim 2-3$ pc
\cite{vandenBergh2010}.  Since massive BBHs should have sunk to the
center due to dynamical friction, the acceleration parameter could
increase to $\epsilon\approx 100~(r/\pc)^{-1}$.  However, many BBHs
would be ejected from the GCs, reducing their acceleration back to
values for orbits in the halo ($\epsilon \lesssim 1$).

LIGO binaries could also be formed in NSCs and/or in galactic nuclei
due to mass segregation through dynamical friction
\cite{O'Leary_2009,Antonini_2016}.  Since the escape velocity from
these systems is higher, a larger number of BBHs can remain within
smaller radii of $r\sim 1$ pc with velocities of $\sim 30-100~\kms$.
The acceleration parameter for these binaries would be larger,
$\epsilon \simeq 10^3~(v/30~\kms)^2(r/\pc)^{-1}$.

\vspace{1mm} \emph{C. Binary BH formation in AGN disks} --- It is
possible to form BBHs, detectable by LIGO, with the help of AGN disks.
They could form either from massive stellar binaries in the disk
itself, at a few pc from the central SMBHs \cite{Stone_2017} or at
migration traps located closer in \cite{Bellovary+2016}; pre-existing
binaries in the 3D bulge can also be captured in the inner regions
($<1~\pc$) of the disk \cite{Bartos_2016}.  In these scenarios, SMBHs
with masses of $10^{6-7}~\msun$ likely dominate, since they are the
most numerous and most efficiently accreting SMBHs with the densest
disks in the local universe~\cite{GreeneHo2007,GreeneHo2009}.  At the
location of the birth of the BBHs ($\sim 1~\pc$), their orbital
velocity around the SMBH would be $\sim 200~\kms$.  The acceleration
is already as high as $\epsilon \simeq 4\times
10^3(M_\bullet/10^7~\msun)(r/1~\pc)^{-2}$.  However, these binaries
are expected to migrate inward through the accretion disk, and many of
them may be located closer to the center when they enter the LISA band
\cite{Bellovary+2016,Bartos_2016,Stone_2017}.  At a distance of
$0.1~\pc$ from the center, the Keplerian velocity increases and the
acceleration parameter is $\epsilon \sim 10^5$.

\vspace{1mm} \emph{D. Very high-z binaries} --- Finally, another
scenario is massive BBH formation in extremely metal-poor environments
at high redshift.  The first generation of stars in the universe at
$z>10-20$, the so-called Population III (PopIII) stars, are typically
as massive as $\sim 10-300~\msun$ \cite{Hirano_2014}.  PopIII binaries
form coalescing BBHs efficiently, which can contribute to the rate of
detectable LIGO events \cite{K14}, including the existing O1
detections \cite{Inayoshi_2017}.
PopIII remnants are expected to be located inside spiral galaxies like
Milky-Way in the current universe
\cite{Tumlinson_2010,Ishiyama_2016,Griffen_2016}.  Cosmological N-body
simulations have suggested that PopIII remnants are distributed in the
bulge, with $\sim 0.1-1\%$ of the remnants concentrated inside
$r\lesssim 1$ kpc. In this case, the acceleration parameter is
$\epsilon \simeq 10-100$.

\vspace{-0.5\baselineskip}
\section{Discussion and implications}
\label{sec:discuss}
\vspace{-0.5\baselineskip}

Different formation scenarios of stellar-mass
BBHs predict a wide range of typical acceleration parameters (see
Table~1).  As pointed out in \cite{Bonvin_2016} and confirmed by our
analysis, in the BBH formation scenarios with low values of $\epsilon
< 100$, the effect of the CoM acceleration of merging BBHs is
difficult to observe by LISA in the operation time of $\delta t \simeq
5$ yr.  On the other hand, BBHs produced in NSCs and in AGN disks are
expected to have large and measurable accelerations, with $\epsilon
>10^3$.  Moreover, the number density of the NSCs ($n_{\rm NSC}\simeq
10^{-2}~\mpc^{-3}$ \cite{Miller_2009}) and AGN 
($n_{\rm AGN}\simeq 10^{-5}~\mpc^{-3}$ \cite{Ueda_2014})
are higher than the number density of merging BBHs at $f \simeq 14$
mHz, $n_{\rm m}(=Rf/\dot{f})\simeq 10^{-6}~\mpc^{-3}$, inferred from the
existing LIGO detections.  Thus, LISA will likely observe phase drifts
in the GW inspiral waveform if these scenarios contribute
significantly to the total event rate.  
%Conversely, a non-detection of
%such phase drift will imply that the NSCs and AGN disk formation
%channels are sub-dominant.

Conversely, if no acceleration is detected among a total of $N_{\rm tot}$ events 
in the region of measurable parameter space,
then this requires that other formation channels, not involving SMBHs are dominant.  
To illustrate this quantitatively, let us consider BBHs with intrinsic acceleration $\epsilon$, 
and suppose that line-of-sight accelerations 
$\epsilon_{\rm los}= |\cos \theta | \epsilon \geq \epsilon_{\rm obs}$ are detectable.  
Assuming that $\theta$, the angle between the line of sight to the BBH and 
the BBH's instantaneous acceleration vector, has an isotropic distribution, 
we would expect $N_{\rm det}=f(\epsilon) N_{\rm tot} [1 - \epsilon_{\rm obs}/\epsilon]$ 
events with measurable acceleration (i.e., within double cones with 
$|\cos \theta | \geq \epsilon_{\rm obs}/\epsilon$), where $f(\epsilon)$ is the fraction of events
with a 3D acceleration above $\epsilon$.  Setting $N_{\rm det}<1$ yields the upper limit 
$f(\epsilon)< N_{\rm tot}^{-1} [1 - \epsilon_{\rm obs}/\epsilon]^{-1}$.  
For example, if we have $N_{\rm tot} = 100$ events and the sensitivity was 
$\epsilon_{\rm obs}=10^4$, then at most $3\%$ of BBHs could have 
$\epsilon > 1.5\times 10^4$, or be located within 
$\lesssim 0.5~\pc~(M_\bullet/10^7{\rm M_\odot})^{1/2}$ from SMBHs.

The sky position and the distance to merging BBHs for $\delta t>2$ yr
with a high SNR ($\rho \gtrsim 10$) can be estimated by LISA alone to
a statistical accuracy of
$\Delta \Omega_{\rm s}\simeq 1.2~f_{14}^{-2}(\rho/10)^{-2}~{\rm deg}^2$ and 
$\Delta d_L/d_L\simeq 0.2(\rho/10)^{-1}$
%
%\begin{equation}
%\Delta \Omega_{\rm s}\simeq 1.2\, {\rm deg}^2
%\left(\frac{f}{14~{\rm mHz}}\right)^{-2}
%\left(\frac{\rho}{10}\right)^{-2},
%\end{equation}
%
%\begin{equation}
%\frac{\Delta d_L}{d_L}\simeq 0.2\left(\frac{\rho}{10}\right)^{-1},
%\end{equation}
%
\cite{Takahashi_Seto_2002,Kyutoku_Seto_2016}. 
The corresponding error volume is given by
$\Delta V=d_L^2\Delta d_L \Delta \Omega 
\simeq 9.6\times 10^{3}
~f_{14}^{-2}(\rho/10)^{-6}~\mpc^3$.
Note that $\rho\simeq 10 (d_L/510~\mpc)^{-1}$.  Since AGN are rare
objects, with abundance of a few $\times 10^{-5}\,{\rm
  Mpc}^3$~\cite{GreeneHo2007,GreeneHo2009}, the number of random
interloping AGN within the error volume is well below unity even for
$\rho=8$.  This means that the AGN hosts can be identified uniquely
from LISA observations alone, without EM counterparts.  
By comparison, the advanced LIGO-Virgo O3 observing run can achieve a 3D error volume
of $\sim 10^5~\mpc^3$ or better only for $<10\%$ of merging BBHs with
$30+30~\msun$ \citep{Chen_Holz_2016}. This still allows a secure
identification of the connection with AGN hosts, but only
statistically \cite{Bartos_2017}.  
In the NSC scenario, the LISA error
volume contains several candidate host galaxies even for relatively
high SNR, $\rho >15$ ($d_L<340~\mpc$), so that one would have to
resort to a statistical correlation between LISA events and NSCs.

The LISA data predict the coalescence time of BBHs with an error of
$<10$ s \cite{Sesana_2016}.  However, this prediction would be biased 
due to the CoM acceleration of the BBHs \cite{Bonvin_2016}.  This bias has to be 
taken into account for any advance planning of follow-up EM
observations of the merging BBHs. The phase drifts caused by the
acceleration could be partially mimicked by a slight change in the mass ratio and time of coalescence. 
However, our Fisher analysis indicates that for sources with
$35~\msun\lesssim M_{\rm cz}\lesssim 63~\msun$ which chirp inside the
LISA band for 2-5 yrs, and especially for those whose eventual merger
is detected by LIGO, these degeneracies are mitigated, and a
measurement of the acceleration remains viable
(see~Figs.~\ref{fig:Mc_fmin_contourplots} and \ref{dL_vs_eps}).  Such
a measurement will robustly test formation channels of coalescing
stellar-mass BBHs involving an SMBH in a galactic nucleus.

The possibility of measuring the CoM acceleration of a merging BBH due
to a nearby SMBH has been previously discussed for extreme mass ratio
inspirals ($10^{5-6}~\msun + 10~\msun$) in the LISA \cite{Yunes_2011b}
band, and for stellar-mass BBHs in the LIGO band \cite{Meiron_2017}.
In the latter case, detection of the phase drift of the BBH during the
handful of orbits executed inside the LIGO band requires an extremely
close separation between the BBH and the SMBH ($\sim 10^{11}\,{\rm
  cm}$); these rare cases of extremely close-in binaries would however
provide the opportunity to measure several other relativistic effects.

%\newpage
\vspace{5mm}
%\vspace{-\baselineskip}

\begin{acknowledgments}
We thank Enrico Barausse, Camille Bonvin, Tomoaki Ishiyama, Antoine
Klein, Nicholas Stone, Riccardo Sturani and Kent Yagi for useful
discussions.  This work is partially supported by the Simons
Foundation through the Simons Society of Fellows (KI) and by a Simons
Fellowship in Theoretical Physics (ZH), and by NASA grant NNX15AB19G
(to ZH).
NT acknowledge support from the Labex P2IO and an Enhanced Eurotalents Fellowship.
\end{acknowledgments}

\vspace{-\baselineskip}

\bibliography{ref}

\end{document}